\documentclass[prl,aps,twocolumn,draft,showpacs]{revtex4}

\begin{document}

\draft
\title{ Fatal Missing Link from Classical Brane Dynamics to Einstein Gravity 
	\\ and a Quantum Theoretical Solution }
\author{			Keiichi Akama}
\address{	Department of Physics, Saitama Medical University,
 			 Saitama, 350-0496, Japan}
\date{\today}

\begin{abstract}
It is argued that the Einstein equation on a braneworld at low curvature cannot be derived from its classical brane dynamics (i. e. dynamics of the brane as a classical object in the higher dimensional spacetime) as far as the action of the system has smooth low-curvature limit. We discuss possible solutions to the problem.
\end{abstract}

\pacs{ 04.50.+h, 04.60.Pp, 11.25.-w, 12.60.Rc}




\maketitle

The general relativity is one of the most celebrated theory today
	both for its theoretical beauties 
	and for the splendid phenomenological successes. 
According to it, the spacetime is curved due to the matters within it.
Owing to Riemannian geometry, the curvature is described entirely 
	from the inside of the spacetime.
Though it is mathematically complete, 
	an intuitive way for us to understand the curvature 
	is to imagine our 3+1 spacetime as a submanifold (or a ``brane")
	embedded in a higher dimensional spacetime,
	in analogy with the curve or surface in our 3 space. 
The 3+1 submanifold where we live 
	is called ``braneworld" in the current terminology.
In fact, many works were performed based on such embedding pictures
	\cite{BW0}.
The old attempts, however, did not really take the brane
	as a dynamical object in the higher dimensions. 
The model of the brane with the Einstein-Hilbert action, unfortunately, 
	turned out not to reproduce the Einstein gravity \cite{RegTei}.
On the other hand, it was shown that 
	the Einstein gravity can be induced by the quantum effects 
	of the 3-brane Lagrangian of the Dirac-Nambu-Goto type \cite {Akama78}.
Dynamical models of the braneworld were constructed 
	with a soliton solution in the higher dimensional spacetime
	\cite {BW1}.
Models of the braneworld were considered from various points of view
	\cite{BW2}.
In the string theory, people drew physical pictures
	where a D-brane is taken as our spacetime \cite{Dbrane}.
Attempts were made to solve the hierarchy problem 
	between the Planck scale and the particle physics scale
	with large extra dimensions\cite{AADD}, 
	or with warped extra dimensions \cite{RS}. 
Possibilities were discussed to observe productions and decays of Kaluza-Klein modes
	and blackholes in particle collisions available in the near future.
It was argued that a warped extra dimension can localize the bulk graviton 
	around the brane.
Various new possibilities were pursued in the braneworld cosmology \cite{cosm}.
Various remarkable aspects of the brane induced gravity were discussed
	\cite{BIG}.

Here we inquire the models on the criterion (A) that
  	{\it the induced metric indicated by brane dynamics
	should obey the Einstein equation on the brane for small curvatures}.
Because the brane is a dynamical object in the higher dimensions,
	it obeys its own equation of motion,
	which entirely determines the developments of the induced metric 
	on the brane. 
Then the criterion (A) implies that
	the Einstein equation 
	should be derived from the equation of motion of the brane
	for small curvatures.
Otherwise we cannot claim that the braneworld predicts 
	what are predicted by the Einstein gravity and are observed.
In principle, it is sufficient if theories satisfy the weaker property 
(B) that  
	{\it  the induced metric indicated by brane dynamics
	should clear the phenomenological requirements 
	cleared by the Einstein equation}.
In practice, however, many models with (B)
	would effectively satisfy Einstein equation for small-curvatures,
	{\it i.\ e.} fulfill the criterion (A).
The idea of brane-world itself is strongly motivated by
	the picture of the curved spacetime in the Einstein theory
	with an expectation that we can inherit 
	its beautiful and successful results.
Thus criterion (A) is natural and useful 
	for the braneworld theories.

The criterion (A) may look obvious and 
	one may expect that it is easily fulfilled.
In fact, specific brane configurations in concern were taken to satisfy
	the brane Einstein equation in many models.
For example, the flat brane, as is adopted in many models,
	satisfies the Einstein equation of the empty brane.
They are, however, insufficient.
Such checks are meaningful only when the Einstein equation
	is required by the brane dynamics.
To show that dynamics require the Einstein equation,
	we need to establish that not only the specific but  	
	all the configurations indicated by the dynamics
	do satisfy the Einstein equation.
In this paper, we inquire models of brane dynamics on the criterion (A), and 
	argue that, unfortunately, {\it the Einstein equation
	is hard to hold on the brane with the classical induced metric 
	for small curvatures}.  
In fact, no successful model of dynamics has been found so far. 
The fundamental link is missing in the braneworld scenario!
Later we will discuss possible cures for the defect. 
Among them, promising is the quantum braneworld, 
	where the Einstein gravity with composite metric 
	is induced through large quantum effects of the brane. 
We will demonstrate remarkable quantum theoretical features of the scheme
	in recovering the missing linkage.

Let us consider $(p-1)$-brane embedded in the $D$-dimensional spacetime
	with the coordinate $X^A$ and 
	the metric tensor $G_{IJ}(X)$ ($I,J,C=0,1,\cdots,D-1$).
Let $Y^I(x^\mu)$ ($I=0,1,\cdots,D-1$) be the position of the brane, 
	where $x^\mu$ ($\mu=0,1,\cdots,p-1$) are parameters.
Then, the induced metric on the brane is given by \cite{comma}
\begin{eqnarray}
	g_{\mu\nu}=Y^I_{,\mu} Y^J_{,\nu} G_{IJ}(Y).
\label{g_munu}
\end{eqnarray}
The brane cannot be specified solely by $g_{\mu\nu}$,  
	but is specified by its position $Y^I(x^\mu)$. 
The brane curvature tensor $R_{\lambda \mu\nu\rho}$ 
	is written in terms of $Y^I$ as \cite{comma}
\begin{eqnarray}
	R_{\lambda \mu\nu \rho }
	&=& (Y^I_{;\lambda \rho } Y^J_{;\mu\nu}
	- Y^I_{;\mu \rho } Y^J_{;\lambda \nu})G_{IJ}(Y)
\cr&& 
	+Y^I_{,\lambda}Y^J_{,\mu} Y^K_{,\nu} Y^L_{,\rho} \hat R_{IJKL}(Y) 
\label{RY}
\end{eqnarray}
with the covariant derivative
\begin{eqnarray}
	Y^I_{;\mu\nu}&=&Y^J_{,\mu\nu}-Y^J_{,\lambda}\Gamma^\lambda_{\mu\nu}
	+ Y^J_{,\mu}Y^K_{,\nu}\hat\Gamma^I_{JK}(Y)
\cr
	&=&(\delta^I_J-Y^I_{,\rho }Y^K_{,\lambda}g^{\rho\lambda}G_{KJ}(Y))
\cr&& \ \ \ \ \ \ \ \ \
	\times (Y^J_{,\mu\nu}+ Y^L_{,\mu}Y^M_{,\nu}\hat\Gamma^J_{LM}(Y)),
\label{covariand derivative}
\end{eqnarray}
	where $\Gamma^\lambda_{\mu\nu}$ is the Christoffel symbol
	in terms of $g_{\mu\nu}$,
	$\hat\Gamma^I_{JK}(Y)$ and $\hat R_{IJKL}(Y)$ are
	the Christoffel symbol and the curvature tensor, respectively, 
	written in terms of $G_{IJ}(Y)$, and
	the inverse $g^{\rho\lambda}$ of $g_{\mu\nu}$
	depends on $Y^I$ according to (\ref{g_munu}). 
Then, the Einstein equation is written in terms of $Y^I$ as \cite{comma}
\begin{eqnarray}&&
	g^{\lambda\rho}
	(\delta_\sigma^\mu \delta_\tau^\nu- g_{\sigma\tau}g^{\mu\nu}/2)
	Y^I_{;\lambda [\rho } Y^J_{;\nu]\mu}G_{IJ}(Y)
\cr&& 
	+g^{\lambda\rho}
	(\delta_\sigma^\mu \delta_\tau^\nu- g_{\sigma\tau}g^{\mu\nu}/2)
	Y^I_{,\lambda} Y^J_{,\mu} Y^K_{,\nu} Y^L_{,\rho} \hat R_{IJKL}(Y)
\cr&& 
	+\lambda Y^I_{,\sigma } Y^J_{,\tau} G_{IJ}(Y)
	=-8\pi G_{\rm N}T_{\sigma\tau},  
\label{EinsteinY}
\end{eqnarray}
where $G_{\rm N}$ and $\lambda$ are the Newtonian gravitational 
	and the cosmological constants, respectively, 
	$T_{\mu\nu}$ is the energy momentum tensor,
	and $g^{\mu\nu}$ depends on $Y^I$ according to (\ref{g_munu}).
But (\ref{EinsteinY}) can not be the equation of motion for $Y^I$, because
	(i) it does not determine $Y^I$ without ambiguities, and 
	(ii) $Y^I$ should have its genuine equation of motion 
	arising from its dynamical setup.
To have Einstein gravity for small curvature region,
	we should derive (\ref{EinsteinY})
	from the equation of motion of the system
	at least within appropriate approximations.

Let us examine examples of dynamics.
The simplest and widely adopted dynamics for the brane is
	that with the Dirac-Nambu-Goto action 
\begin{eqnarray}
	\xi\int\sqrt{-g}d^{p} x,
\label{DNG}
\end{eqnarray}
	where $g=\det g_{\mu\nu}$ is taken to be written in terms of $Y^I$
	according to (\ref{g_munu}), and $\xi$ is a constant.
This is expected to take place in the low curvature limit of various models.
Variation of (\ref{DNG}) with respect to $Y^I$ leads to the equation of motion 
\begin{eqnarray}
	g^{\mu\nu}Y^I_{;\mu\nu}=0.
\label{DNG_EM}
\end{eqnarray}
Unfortunately, we cannot derive the Einstein equation (\ref{EinsteinY}) 
	from the brane equation of motion (\ref{DNG_EM}).
It is established by finding a counter example. 
For example, we can show the following type of solutions of (\ref{DNG_EM}) 
	does not satisfy (\ref{EinsteinY}) 
	up to $O(a^3)$,
\begin{eqnarray}
\cases{ 
	Y^\mu=x^\mu \mbox{ for } \mu< p,
\cr	Y^{p}=a/\sqrt{(x^1)^2+(x^2)^2+(x^3)^2}+O(a^3),
\cr	Y^m=0 \mbox{ for } m > p,}
\label{DNGsolution}
\end{eqnarray}
	where $a$ is a constant such as $\hat R_{IJKL}\sim O(a^2)$.

Another example of the brane dynamics is 
	that with the Einstein-Hilbert action written in terms of $Y^I$:
\begin{eqnarray}
	\kappa\int\sqrt{-g}Rd^{p} x,
\label{RT}
\end{eqnarray}
	where $R=g^{\mu\nu}R_{\mu\nu}$, 
	$R_{\mu\nu}=g^{\rho\sigma}R_{\rho\mu\nu\sigma }$
	and $\kappa$ is a constant.
Variation of (\ref{RT}) with respect to $Y^I$ leads to the equation of motion 
\begin{eqnarray}
	\left(R^{\mu\nu}-{1\over2}Rg^{\mu\nu}\right)Y^I_{;\mu\nu}=0. 
\label{RTEOM}
\end{eqnarray}
As was pointed out in \cite{RegTei} long ago,
	the equation (\ref{RTEOM}) does not imply the Einstein equation 
$ 	R^{\mu\nu}- R g^{\mu\nu} /2=0 $
	for this system, 
	because the multiplied tensor $ Y^I_{;\mu\nu}$ in (\ref{RTEOM})
	obeys the identity 
$	Y_{I,\lambda}Y^I_{;\mu\nu}=0. $ 
Many solutions of (\ref{RTEOM}) does not satisfy 
	the Einstein equation.

Many models of the braneworld were considered 
	based on 3-dimensional solitonic objects in higher dimensions,
	including D-branes and spacetime singularities.
In any case, we can extract the position variable $Y^I(x^\mu)$, 
	as far as it is interpreted as a brane.
In general, it would be difficult to work out precise effective dynamics 
	for $Y^I(x^\mu)$ imposed by their physical setup.
For small curvatures, however, it reduces 
	to the Dirac-Nambu-Goto action (\ref{DNG})
	or, in the next-to-leading order, 
	to the Einstein-Hilbert action (\ref{RT}),
	as far as the action has regular small curvature limit.
As is discussed above,  both of them do not lead to the Einstein equation.
This suggests that the Einstein equation is hard to be reached
	for small curvatures on the classical braneworld.
In fact, no successful model is found so far.

Now we discuss possible ways to recover this serious missing link
	in the braneworld scenario.
The reason why the Einstein equation (\ref{EinsteinY}) by itself 
	cannot be the equation of motion of the brane is that
	it cannot uniquely specify the motion of the brane position $Y^I(x)$.
A way out of this situation is
	to impose the Einstein equation (\ref{EinsteinY})
	together with the equations of motion on
	the extrinsic curvature $B^a_{\mu\nu}$
	and the normal connection field $A^{ab}_\mu$ defined by
\begin{eqnarray}
	B^a_{\mu\nu}=(n^a)^I Y^J_{;\mu\nu}G_{IJ},\ \ \       
	A^{ab}_\mu= (n^a)^I (n^b) ^J_{;\mu} G_{IJ},    \label{A}
\end{eqnarray}
where $ n^a $ $(a=p,\cdots,D-1)$ is the orthonormal system 
	of the normal vectors of the brane at each point on the brane.
These fields are not all independent 
	and should obey the Gauss-Codazzi-Ricci equations  \cite{comma}
\begin{eqnarray}
	\Xi_{\kappa\lambda \mu\nu}
	&\equiv &R_{\kappa\lambda \mu\nu}
	+B_{a\kappa[\mu}B^a_{ \nu]\lambda }
\cr	&&-Y^K_{,\kappa} Y^L_{,\lambda} Y^M_{,\mu} Y^N_{,\nu} \hat R_{KLMN}=0,      
\cr	\Xi_{ a\lambda \mu\nu}    
	&\equiv &B_{a\lambda [\mu,\nu]}
	+A_{ab[\mu}B^b_{ \nu]\lambda }
\cr	&&-(n _a)^K Y^L_{,\lambda} Y^M_{,\mu} Y^N_{,\nu} \hat R_{KLMN}=0,      
\cr	\Xi_{ ab \mu\nu}
	&\equiv &A_{ab [\mu,\nu]}
	+A_{ac[\mu}A^c{}_{ b \nu] }
	+B_{a\rho[\mu}B_b{}^{\rho}{}_{ \nu] }
\cr	&&-(n _a)^K (n _b)^L Y^M_{,\mu} Y^N_{,\nu} \hat R_{KLMN}=0     
    \label{GCR}
\end{eqnarray}
for $\kappa,\lambda,\mu,\nu=0,\cdots,p-1$ and $ a,b,c=p,\cdots,D-1$.
The fundamental theorem of hypersurface guarantees
	unique existence of the system of $Y^I$ 
	up to its initial point values
	for a given system of $g_{\mu\nu}$,
	$B_{a\mu\nu}$ and $A_{ab\mu}$ satisfying (\ref{GCR}).
It is, however, too restrictive to impose the equations of motion
	and the Gauss-Codazzi-Ricci equations simultaneously.
Too many equations for the number of the variables.
A dynamically consistent procedure is 
	to introduce Lagrange multiplier $u^{KL\mu\nu}$ 
	for the Gauss-Codazzi-Ricci equations so that
\begin{eqnarray}
	\int \sqrt{-g} 
	({\cal L}+u^{KL\mu\nu}\Xi_{KL\mu\nu})
	d^p x,
\label{Lagmul}
\end{eqnarray}
where $\cal L $ is the Lagrangian which leads 
	to the above-mentioned equations of motion
	for the variation of the fields $g_{\mu\nu}$, $B_{a\mu\nu}$ and $A_{ab\mu}$.
In this way we can describe dynamics of branes 
	respecting the Einstein equation
	with some modifications due to the Lagrange multiplier.

This scheme, however, suffers from the following restriction.
The equations of motion are derived 
	via variations with respect to 
	$g_{\mu\nu}$, $B^a_{\mu\nu}$ and $A^{ab}_\mu$,
	but not to $Y^I$.
Since $Y^I$ is obtained by integrating the Gauss-Weingarten equation,
	it depends on $g_{\mu\nu}$, $B^a_{\mu\nu}$ and $A^{ab}_\mu$
	non-locally in $x^\mu$.
Therefore, small variations in $g_{\mu\nu}$, $B^a_{\mu\nu}$ and $A^{ab}_\mu$
	cause in general large variation in $Y^I$.
In order that the variation principle with respect to
	$g_{\mu\nu}$, $B^a_{\mu\nu}$ and $A^{ab}_\mu$ makes sense,
	the action cannot explicitly depend on $Y^I$.
This is not desirable because explicit dependence on $Y^I$ 
	is necessary to describe the local interactions of the brane
	with bulk objects or other branes.
Strictly speaking, the $Y^I$-dependence of $\hat R_{KL\mu\nu}$
	in $\Xi_{KL\mu\nu}$ in (\ref{Lagmul}) can not get rid of this problem.
Hence, this scheme is applicable only to
	non-interacting single brane embedded 
	in a higher spacetime with constant curvature.

A more natural and promising solution for the missing link problem is 
	given by taking into accounts the quantum effects.
A troublesome obstacle in the classical brane 
	is the deterministic relation (\ref{g_munu})
	between $g_{\mu\nu}$ and $Y^I$. 
Owing to it, the dynamics of $Y^I$ entirely determines the dynamics of $g_{\mu\nu}$,
	which differs from the Einstein equation.
In quantum theory, however, the operator products in (\ref{g_munu})
	at short distances suffer from quantum corrections. 
If we introduce auxiliary field for metric tensor with the constraint (\ref{g_munu}),
	the quantum corrections depends on them.
The quantum effects are divergent and not renormalizable
	if they persist down to the null distance.
In practice, we expect that the brane has thickness $\delta$.
Then, the effects concerned with the metric should be cutoff at $\delta$.
Dimensional analyses show that the leading contribution is proportional 
	$\delta^{-p}g_{\mu\nu}$, the cosmological term, 
	and the next to leading contribution is proportional to
	$\delta^{-p+2}(R_{\mu\nu}-g_{\mu\nu}R/2)$, the Einstein tensor. 
Thus the auxiliary field metric obtains its kinetic term,
	and gets an independent degrees of freedom.
Namely the composite graviton is induced \cite{indg}.
The emergence of such quantum composite fields 
	is seen in a wide class of field theories,
	and has been extensively studied \cite{comp}.
The right-hand side of (\ref{g_munu}) can be interpreted as
	the energy momentum tensor of the ``field" $Y^I$ on the brane.
Thus the troublesome relation (\ref{g_munu}) in the classical braneworld
	is no longer true, and instead we have the Einstein equation
	with cosmological term of $O(\delta^{-p})$, 
	the Newtonian constant of $O(\delta^{p-2})$,
	and smaller correction terms of $O(R^2)$.
The cosmological term should be set small by fine tuning,
	and the thickness $\delta$ should be of the order of the Planck scale.

Now we see the effects more precisely in a simple model.
We start with the Dirac-Nambu-Goto action (\ref{DNG}).
It is equivalent to 
\begin{eqnarray}
	\xi\int\sqrt{- g}	\left(
	g^{\mu\nu}Y^I_{,\mu} Y^J_{,\nu}G_{IJ}(Y)-p+2
	\right) d^{p} x,
\label{DNGaux}
\end{eqnarray}
	where $ g_{\mu\nu}$ and $ g=\det g_{\mu\nu}$
	are taken as independent variables without (\ref{g_munu}).
In (\ref{DNGaux}), the  $ g_{\mu\nu}$ has no kinetic term,
	and is an auxiliary field.
Instead, variation of (\ref{DNGaux}) with respect to $g_{\mu\nu}$
	automatically reproduces the constraint (\ref{g_munu}).
The actions (\ref{DNG}) and (\ref{DNGaux}) are equivalent in quantum level,
	because the equations of motion, 
	as well as the Dirac bracket algebra coincide \cite{Akama:1979tm}.
Now we consider the quantum fluctuations of the ``field" $Y^A(x^\mu)$
	on the auxiliary field $g_{\mu\nu}(x^\mu)$.
Since the divergent contributions stated above are dominated by
	the short-distance effects of $O(\delta)$,
	we neglect the small curvature of the bulk spacetime in (\ref{DNGaux}).
We expand the action (\ref{DNGaux}) in terms of 
	the quantum fluctuations $\varphi^m=Y^m$ $(m\ge p)$ 
	around the classical solution $Y^\mu=x^\mu$ $(\mu< p)$, $Y^m=0$ $(m\ge p)$.
Then, we calculate the one-loop diagrams with internal $\varphi$-lines. 
Retaining the leading and the next-to-leading parts in $1/\delta$, 
	we obtain the following effective action 
	for the auxiliary field $g_{\mu\nu}$:
\begin{eqnarray}
	{D-p \over (2\sqrt{\pi}\delta) ^p}
	\int\sqrt{- g}	\left(
	{1\over p} +  {\delta^2\over6(p-2)}R
	\right) d^p x,
\label{S_eff}
\end{eqnarray}
where we have used the heat kernel regularization 
	with a cutoff of the Schwinger proper time at $\delta$.
The induced Einstein-Hilbert action in (\ref{S_eff}) has positive sign,
	which indicates that the induced gravity is attractive as desired.
The induced cosmological term in (\ref{S_eff}) has also positive sign.
To have the correct sign for the kinetic term of $\varphi^m$ in (\ref{DNGaux})
	we should take $\xi<0$, 
	which means that the cosmological term arising from (\ref{DNGaux})
	is also positive.
These signs are not desired for cancellation.
We expect some other effects such as fermionic fluctuations to cancel them
	in the more realistic models.

Thus we expect that the serious missing link 
	in the classical braneworlds to Einstein gravity 
	would be recovered in the quantum brane 
	through its large quantum effects at short distances.
It is remarkable if the gravitation so familiar to us 
	is originated from the quantum fluctuations of our thin spacetime itself.

We would like to thank 
Professor E.~J.~Copeland, 
Professor G.~Dvali,
Professor G.~Gabadadze,
Professor G.~Gibbons, 
Professor T.~Hattori,
Professor P.~Kanti, 
Professor S.~Randjbar-Daemi, and 
Professor \\M.~Shaposhnikov 
for discussions.


\begin{thebibliography}{99}

\bibitem{BW0}
C.~Fronsdal, Nuovo Cim.\ {\bf 13}, 5 (1959);
D.~W.~Joseph, Phys.\ Rev.\ {\bf 126}, 319 (1962).

\bibitem{RegTei} 
T.Regge and C.Teitelboim, 
in {\it Marcel Grossman Meeting on Relativity, 1975}(North Holland, 1977) 77.

\bibitem{Akama78} 
K.~Akama,  {Prog.\ Theor.\ Phys.} {\bf 60}, 1900 (1978).

\bibitem{BW1}
K.~Akama, Lect.\ Notes in Phys.\ {\bf 176}, 267 (1983);
V.~A.~Rubakov and M.\ E.\ Shaposhnikov,
Phys.\ Lett.\ {\bf B125}, 136 (1983).

\bibitem{BW2}
M.~D.~Maia, J.\ Math.\ Phys.\ {\bf 25}, 2090 (1984); 
M.~Visser, Phys.\ Lett.\ {\bf B159}, 22 (1985);
M.~Pav\v si\v c, Class.\ Quant.\ Grav.\ {\bf 2}, 869 (1985);
G. W.  Gibbons and D. L. Wiltshire, Nucl.\ Phys.\ {\bf B287}, 717(1987); 
K.~Akama,  Prog.\ Theor.\ Phys.\ 
{\bf 78}, 184 (1987); {\bf 79}, 1299 (1988); {\bf 80}, 935 (1988).

\bibitem{Dbrane}
J.~Polchinski, Phys.\ Rev.\ Lett.\ {\bf 75}, 4724--4727 (1995);
P.~Horava and E.~Witten, 
Nucl.\ Phys.\ {\bf B460}, 506--524 (1996); Nucl.\ Phys.\ {\bf B475}, 94--114 (1996).

\bibitem{AADD} 
N.~Arkani-Hamed, S.~Dimopoulos, and G.~Dvali, 
Phys.\ Lett.\ {\bf B429}, 263-272 (1998); 
Phys.\ Rev.\ {\bf D59}, 086004 (1999); 
I.~Antoniadis, N.~Arkani-Hamed, S.~Dimopoulos, and G.~Dvali, 
Phys.\ Lett.\ {\bf B436}, 257-263 (1998).

\bibitem{RS} 
L.~Randall, R.~Sundrum, 
Phys.\ Rev.\ Lett.\ {\bf 83}, 3370-3373 (1999); 4690-4693 (1999).

\bibitem{cosm}
  P.~Kanti, I.~I.~Kogan, K.~A.~Olive and M.~Pospelov,
  Phys.\ Lett.\ B {\bf 468}, 31 (1999);
T.\ Shiromizu, K.\ I.\ Maeda, and M.\ Sasaki, Phys.\ Rev.\ {\bf D62}, 024012 (2000); 
  R.~Maartens, D.~Wands, B.~A.~Bassett and I.P.C.~Heard,
  Phys.\ Rev.\ D {\bf 62}, 041301(R) (2000);
  P.~Binetruy, C.~Deffayet and D.~Langlois,
  Nucl.\ Phys.\ B {\bf 565}, 269 (2000);
  R.~Gregory, V.~A.~Rubakov and S.~M.~Sibiryakov,
  Phys.\ Rev.\ Lett.\  {\bf 84}, 5928 (2000);
  E.~J.~Copeland, A.~R.~Liddle and J.~E.~Lidsey,
  Phys.\ Rev.\ D {\bf 64}, 023509 (2001);
  R.~Maartens,
  Living Rev.\ Rel.\  {\bf 7}, 7 (2004).


\bibitem{BIG} 
G.\ Dvali, G.\ Gabadadze, and M.\ Porrati, 
Phys.\ Lett.\ {\bf B485}, 208 (2000); 
K.~Akama and T.\ Hattori, Mod.\ Phys.\ Lett.\ {\bf A15}, 2017 (2000);
G.~Dvali and G.~Gabadadze,
Phys.\ Rev.\ D {\bf 63}, 065007 (2001);
C. Charmousis, R. Gregory, N. Kaloper, and A. Padilla, hep-th/0604086 (2006).

\bibitem{comma}
A comma (semicolon) in the suffix array indicates 
(covariant) differentiation 
with respect to the coordinate components indicated by 
the suffices following it.
A suffix with $_[$ on its left side and 
a suffix with $_]$ on its right side are taken as anti-symmetrized. 
\bibitem{indg}
A.~D.~Sakharov,  Dokl.\ Akad.\ Nauk SSSR {\bf 177}, 70 (1967)
	[{Sov.\ Phys.\ Dokl.} {\bf 12}, 1040 (1968)];
K.~Akama, Y.~Chikashige and T.~Matsuki,
Prog.\ Theor.\ Phys.\  {\bf 59}, 653 (1978);
K.~Akama, Y.~Chikashige, T.~Matsuki and H.~Terazawa,
	 {Prog.\ Theor.\ Phys.} {\bf 60}, 868 (1978);
A.~Zee, Phys.\ Rev.\ Lett.\ {\bf 42}, 417 (1979);
S.~L.~Adler, Phys.\ Rev.\ Lett.\ {\bf 44}, 1567 (1980). 

\bibitem{comp}
B.~Jouvet, Nuovo Cim. {\bf 5} (1956) 1133; 
Y.\ Nambu and G.\ Jona-Lasinio,	 {Phys.\ Rev.} {\bf 122} (1961) 345;
J.~D.~Bjorken,                  Ann.\ Phys.\ {\bf 24} (1963) 174;
H.~Terazawa, Y.\ Chikashige and K.~Akama, Phys.\ Rev.\ {\bf D15}, 480 (1977);
K.~Akama, Phys.\ Rev.\ Lett.\  {\bf 76}, 184 (1996);
K.~Akama and T.~Hattori, Phys.\ Rev.\ Lett.\  {\bf 93}, 211602 (2004).

\bibitem{Akama:1979tm}
  K.~Akama,
  Prog.\ Theor.\ Phys.\  {\bf 61}, 687 (1979).


\end{thebibliography}
\end{document}